\documentclass[aps,pra,superscriptaddress,showpacs,showkeys,a4paper,
               10pt,notitlepage,twocolumn]{revtex4-1}
\usepackage[utf8]{inputenc}
\usepackage[english]{babel}
\usepackage{amsmath}
\usepackage{amssymb}
\usepackage{amsfonts}
\usepackage{graphicx}
\usepackage{cancel}
\usepackage{upgreek}
\usepackage{mathtools}
\usepackage[T1]{fontenc}

\usepackage{xcolor}
\definecolor{link_blue}{RGB}{52,46,157}

\usepackage{longtable}
\usepackage[colorlinks,
            citecolor=link_blue,
            linkcolor=link_blue,
            urlcolor=link_blue]{hyperref}
\usepackage[tiny,center,uppercase]{titlesec}

\renewcommand{\vec}{\boldsymbol}

\newcommand\ee{\mathrm{e}}

\DeclareMathOperator{\Ei}{Ei}

\begin{document}

\title{Superradiant parametric X-ray emission}

\author{I.\ D.\ Feranchuk}
\email[Corresponding author: ]{iferanchuk@gmail.com}

\affiliation{Atomicus GmbH, Schoemperlen Str. 12a, 76185 Karlsruhe,
  Germany}

\author{N.\ Q.\ San}

\affiliation{Department of Physics, Faculty of Electricity and
  Electronics, Nha Trang University, Nha Trang, Vietnam}

\author{O.\ D.\ Skoromnik}
\email{olegskor@gmail.com}

\begin{abstract}
  We compute a spectrum of parametric X-ray radiation (PXR) inside a
  crystal from a bunch of electrons, which is periodically modulated
  in density. We consider that the bunch of electrons is exiting from
  a XFEL channel. We demonstrate that in the case of a resonance
  between the frequency of parametric X-ray radiation and a frequency
  of modulation of an electron bunch the sequence of strong
  quasi-monochromatic X-ray pulses is formed --- superradiant
  parametric X-ray emission (SPXE) with frequencies multiples of the
  modulation frequency. The number of photons in the impulse of SPXE
  in the case of an extremely asymmetric diffraction is comparable
  with the photon number in the impulse of a XFEL. Moreover the SPXE
  is directed under the large angle to the electron velocity and every
  harmonic in the spectrum is emitted under its own angle.
\end{abstract}

\maketitle

\section{Introduction}
\label{sec:introduction}

Parametric X-ray radiation (PXR) is a well known mechanism of
radiation from charged particles propagating in a periodic medium
\cite{PXR_Book_Feranchuk, PhysRevAccelBeams.21.014701,
  HAYAKAWA2006102, Lauth2006, Brenzinger1997, PhysRevLett.79.2462,
  rullhusen1998novel, PhysRevAccelBeams.25.040704}. Its qualitative
properties are the emission of quasi-monochromatic X-ray beams under
the large angle to the electron velocity and the possibility to
continuously tune the frequency of the radiation by simply rotating a
crystal. As was firstly demonstrated in the work
\cite{BARYSHEVSKY1984141}, when the electron density in the bunch
reaches a critical value, the parametric beam instability can arise in
analogy with the self amplified spontaneous emission (SASE) in the
undulator of an XFEL. This process leads to a spatial modulation of an
electron beam and the generation of a coherent X-ray radiation under
the scale of the crystal length, which is much smaller than the
typical XFEL undulator lengths. However, under the currently
experimentally accessible electron-beams densities the length under
which the parametric instability can be achieved is substantially
larger than the X-ray absorption length in the crystal. Therefore,
when an electron beam initially does not contain any modulation it is
almost impossible to realize the SASE mechanism inside a crystal.

Before discussing the superradiant parametric X-ray emission (SPXE),
let us briefly revise how the undulator is used in the XFEL case
\cite{RevModPhys.91.035003}. On the one hand due to the SASE mechanism
the electron bunch, which density was initially uniformly distributed,
is transformed into a sequence of micro-bunches that leads to the
spatial modulation of the electron density with the period
$d_{0}$. This period is directly defined by the period of the
undulator. Later when the bunch propagates inside the undulator the
coherent radiation is formed (superradiant emission
\cite{RevModPhys.91.035003}) with the frequency
$\omega_{0} = 2 \pi / d_{0}; \ (\hbar = c = 1)$ and with the intensity
proportional to the square of the number of electrons in the
bunch. The coherent radiation is propagated in a small cone along the
electron velocity. The coincidence (resonance) between the modulation
frequency of the bunch and the frequency of the emitted photons
happens automatically since the beam modulation and the emission
frequency are determined by the same undulator radiation mechanism.

\begin{figure*}[t]
\centering
\includegraphics[scale=1.2]{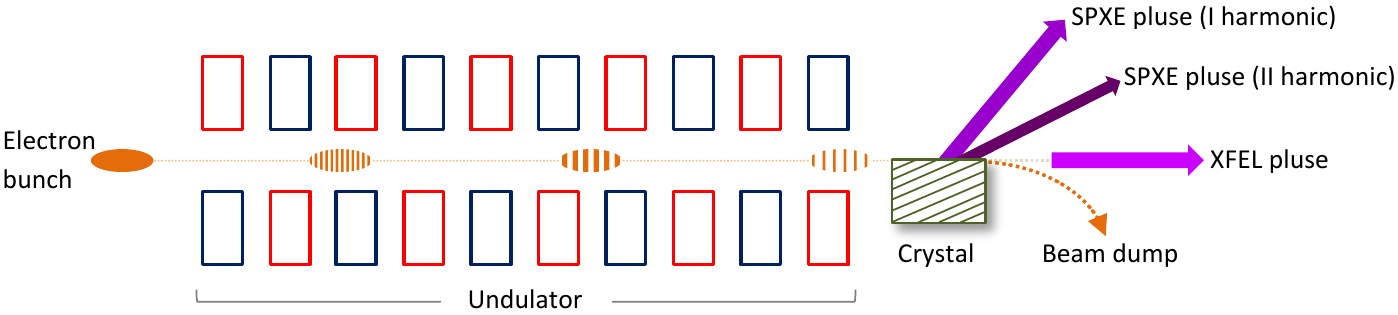}
\caption{Qualitative scheme of the processes which lead to the
  generation of SPXE pulses in the XFEL channel}
\label{fig:1}
\end{figure*}

Now let us consider the SPXE case when we suppose that an electron
beam becomes modulated in density inside the undulator and in the end
enters the crystalline target, where the PXR is generated with the
frequency $\omega_{\mathrm{B}}$ dependent on the crystal structure and
the angle $\theta_{\mathrm{B}}$ between the crystallographic planes
and the electron velocity. As was recently demonstrated in the work
\cite{SKOROMNIK201786} the highest intensity of PXR is reached when
the electrons propagate in the grazing geometry, i.e., in a thin layer
inside the crystal parallel to the crystal-vacuum interface and the
X-ray photons are emitted under the large angle
$2 \theta_{\mathrm{B}}$ to the electron velocity (PXR-EAD). The angle
$\theta_{\mathrm{B}}$ can be chosen in such a way that the resonant
condition $\omega_{0} \approx \omega_{\mathrm{B}}$ is fulfilled. As a
result, in addition to the main XFEL pulse a generation of SPXE will
happen with the intensity also proportional to the square of the
number of electrons in the bunch. According to
Ref.~\cite{PXR_Book_Feranchuk} the spectral density of PXR photons
emitted by a single electron can be larger than the corresponding
density of the undulator radiation. Consequently, the number of SPXE
photons can exceed the corresponding number of the XFEL ones.
Besides, the SPXE photon pulse is directed under the large angle to
the electron velocity, which enlarge the applicability of the XFEL by
the creation of the additional exiting channels of X-rays.
Fig.~\ref{fig:1} presents the qualitative picture of processes which
lead to the SPXE pulses in the XFEL channel.

In our work we describe a generation mechanism of SPXE and
theoretically study characteristics of this new type of the coherent
X-ray radiation.

\section{Qualitative analysis}
\label{sec:qualitative-analysis}

We start the analysis of SPXE from its qualitative estimation by means
of a simple and effective method of the description of electromagnetic
processes of relativistic charged particles interacting with the
medium, namely the method of equivalent photons (pseudo-photons,
Weizsacker-Williams approximation)
\cite{AkhiezerB1965quantum,ter1972high,LandauClassFields}. This
approach is based on the observation that the self field of a
relativistic charged particle is equivalent in its characteristics to
the beam of pseudo-photons with a spectral-angular distribution
$n(\vec k)$ and a narrow angular divergence $\approx \gamma^{-1}$,
which is determined by the relativistic gamma factor of the particle
$\gamma = E/m$ \footnote{Here and through the manuscript we use
  natural units with $\hbar = c = 1$}. As a result, the differential
cross section $d\sigma^{\mathrm{e}}_{if}$ of a transition
$i \Rightarrow f$ between the initial $i$ and final state $f$ of a
charged particle moving with a velocity $\vec v$ and interacting with
the medium is represented as
\begin{align}
  \label{eq:1}
  d\sigma^{\mathrm{e}}_{if} =  n( \vec k) d\sigma^{\mathrm{ph}}_{if}
  (\omega,\vec k_{\perp}) d\omega d\vec k_{\perp},
\end{align}
where $d\sigma^{\mathrm{ph}}_{if}$ is the cross section of the same
transition for a photon with the frequency $\omega$ and the wave
vector $\vec k = (\omega \vec v/v, \vec k_{\perp}) $.

\begin{figure*}[t]
\centering
\includegraphics[scale=1.8]{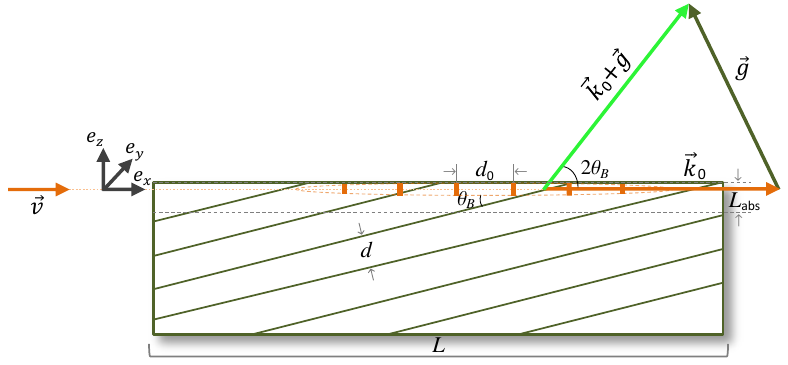}
\caption{Scheme of the photon generation in the case of PXR-EAD geometry}
\label{fig:2}
\end{figure*}

In this framework PXR can be considered as a diffraction of a beam of
pseudo-photons on the crystallographic planes. The spectral-angular
distribution $n( \vec k)$ of pseudo-photons for a single charged
particle is well known
\cite{AkhiezerB1965quantum,ter1972high,LandauClassFields} and is given
by a smooth function of a frequency of the pseudo-photons. For
relativistic particles the wave vector $\vec k$ of a pseudo-photon can
be approximated as $\vec k \approx \vec k_0 = \omega \vec v/v$ with
$|\vec k_{\perp}| \ll k_0$. As a result, PXR peaks are determined by
the frequencies $\omega_{\mathrm{B}}$ when the wave vector
$\vec k_{0}$ satisfies Bragg's condition. Consequently, the emitted
photons are propagating in the directions of $\vec k_{0} + \vec g$,
where $\vec g$ is one of the reciprocal lattice vectors of a
crystal. As was demonstrated in Ref.~\cite{SKOROMNIK201786} PXR will
have the highest intensity when an electron moves in a crystal in a
thin layer near to the crystal surface, EAD geometry, and the emitted
radiation can exit the crystal without absorption, see
Fig.~\ref{fig:2}.

When we now consider an electron bunch consisting of $N$ electrons
uniformly spread in space the total field will be given by an
incoherent sum of fields from each electron. Therefore, the
spectral-angular distribution of the pseudo-photons is simply defined
by the sum of contributions from each particle and equals
$N n(\vec k)$.

However, if the electron beam was moving through the undulator due to
the SASE mechanism it became modulated. Accordingly its density
becomes a periodic function of the longitudinal coordinate with the
period $d_{0}$. Consequently, this suggests that we might expect the
coherent summation of individual fields from every electron. This
results in the substantial modification of total spectral distribution
of pseudo-photons in which we expect to see peaks with the amplitude
proportional to the square of the number of particles in the beam.
The frequencies of these spikes are harmonics of the frequency
$\omega_{0}$. If the frequency of one of the peaks coincides with the
frequency $\omega_{\mathrm{B}}$ we should expect to see the resonant
increase of the intensity of diffracted pseudo-photons. This
corresponds exactly to the SPXE impulse. The SPXE emission happens
when the Bragg's condition $2k_{0} d \sin \theta_B = 2\pi$ for the
reflection of pseudo-photons from the crystallographic planes is
simultaneously fulfilled with the coherence condition of the radiation
from electrons of different micro-bunches $k_{0} d_0 = 2 \pi$, that is
\begin{align}
  \label{eq:2}
   2 d \sin \theta_B = d_0
\end{align}
see Fig.~\ref{fig:2}.

Let us now investigate this process in more details. In the range of
frequencies, which are much smaller than the particle energy $E$,
i.e., $\omega \ll E$ the spectrum of pseudo-photons can be obtained
via classical description \cite{AkhiezerB1965quantum}. The vector
$\vec A(\vec r, t)$ and scalar $\varphi(\vec r, t)$ potentials from a
beam of particles with charge $e_{0}$ that move uniformly in vacuum
are defined by the Maxwell equations
\begin{align}
  \Box \vec A
  &= - 4\pi e_0 \sum_a^N \vec v_a \delta (\vec r - \vec v_a t - \vec r_a), \label{eq:3}
  \\
  \Box \varphi
  &= - 4 \pi e_0 \sum_a^N   \delta (\vec r - \vec v_a t - \vec r_a), \label{eq:4}
\end{align}
where the sum runs over all particles and each particle is located at
the initial position $\vec r_{a}$ and has the velocity $\vec v_{a}$.

The Fourier transform of Eqs.~(\ref{eq:3}-\ref{eq:4}) allows one to
find the potentials and to calculate the electromagnetic fields.

\begin{align}
  \vec  E  (\vec r, t)
  &= \sum_a^N \vec E_a (\vec r, t), \label{eq:5}
  \\
  \vec E_a (\vec r, t)
  &= - \frac{i e_0}{ 2\pi^2} \int d\vec k  \frac{\vec k -
  \vec v_a (\vec k \vec v_a)}{k^2 - (\vec k \vec v_a)^2}e^{i \vec k
  (\vec r - \vec v_a t - \vec r_a)}, \label{eq:6}
  \\
  \vec  H  (\vec r, t)
  &= \sum_a^N \vec v_a \times \vec E_a (\vec r, t). \label{eq:7}
\end{align}

Now we estimate the fields in Eq.~(\ref{eq:5}). For this we consider
that the electron beam has a small angular divergence, such that the
velocity of each particle can be presented as
\begin{align}
  \vec v_a
  &= \vec v + \vec v'_a, \quad  v'_a \ll v, \label{eq:8}
  \\
  1-v^2
  &= \frac{m^2}{E^2} \equiv \gamma^{-2} \ll 1. \label{eq:9}
\end{align}
We direct the $x$ axis along the mean velocity $\vec v$ of the bunch
and suppose that the particle angular divergence is small with respect
to this axis, i.e., $\theta_{a} < \gamma^{-1}$, where the angle
$\theta_{a}$ defines this angular divergence. The fluctuations of the
absolute value of the velocity are related to the nonmonochromaticity
$\Delta E$ of the beam
$|\vec v_a - \vec v| \approx \gamma^{-2}\Delta E/E \ll\gamma^{-2}$
. Taking this into account we can represent the components of the
fluctuation vector $\vec v'_{a}$ of the particle velocity with the
accuracy up t o $\gamma^{-2}$ as
\begin{align}
  \vec  v'_a
  &= \vec \theta_a   - \frac{\theta_a^2}{2}\vec e_{x},
  \quad
  \theta_a^2 = \theta^2_{az}+ \theta^2_{ay}, \label{eq:10}
  \\
  \vec \theta_a
  &= \theta_{az} \vec e_{z} + \theta_{ay} \vec e_{y}, \label{eq:11}
\end{align}
where $\vec e_{x}, \vec e_{y}, \vec e_{z}$ are the unit vectors  (see
Fig.~\ref{fig:2}) .

In this approximation the denominator of the field amplitude in
Eq.~(\ref{eq:6}) equals to
\begin{align}
  k_x^2  \gamma^{-2}   + (\vec k_{\bot} - k_x\vec \theta_a)^2, \label{eq:12}
\end{align}
and demonstrates that the main contribution to the field amplitude
comes from the values
\begin{align}
  \theta_a \simeq \frac{k_{\bot}}{k_x} \simeq \gamma^{-1}. \label{eq:13}
\end{align}
For these angles the field up to accuracy $\gamma^{-1}$ becomes
transverse since $|E_x| \simeq \gamma^{-1} |E_{\bot}|$ and reads
\begin{align}
  \vec  E_{\bot}  (\vec r, t)
  &= \sum_a^N \vec E_{a\bot} (\vec r, t), \label{eq:14}
  \\
  \vec E_{a\bot} (\vec r, t)
  &= - \frac{i e_0}{ 2\pi^2} \int d\vec k
  \frac{(\vec k_{\bot} - \vec \theta_a k_x)e^{i \vec k (\vec r - \vec
  v_a t - \vec r_a)}}{ k_x^2  \gamma^{-2} +  (\vec k_{\bot} - k_x \vec
    \theta_a)^2 }, \nonumber
  \\
  \vec  H  (\vec r, t)
  &=\vec v \times\vec E_{\bot}(\vec r, t). \label{eq:15}
\end{align}

The projection of the energy flux of the electromagnetic field on the
$x$ axis is determined by the following expression
\cite{AkhiezerB1965quantum}
\begin{widetext}
  \begin{align}
    \Pi
    &= \frac{1}{4\pi}\int_{-\infty}^{ \infty}dz dy dt [\vec E \vec
    H]_x = \int_{-\infty}^{ \infty}dx dy dt |\vec E_{\bot}|^2 \approx
    \nonumber
    \\
    &\approx\frac{e_0^2}{2\pi^2 v}\sum_a\sum_b\int d\vec k
      \frac{
      (\vec k_{\bot} - \vec \theta_a k_x)(\vec k_{\bot} - \vec\theta_b k_x)
      e^{i \vec k (\vec r_b - \vec r_a)}e^{i x \vec k( \vec v'_a  - \vec v'_b)}
      }
      {
      (k_x^2\gamma^{-2} + (\vec k_{\bot} - k_x\vec \theta_a)^2)( k_x^2
      \gamma^{-2} + (\vec k_{\bot} - k_x\vec \theta_b)^2)
      }, \label{eq:16}
  \end{align}
\end{widetext}
which can be split into the sum of two parts
\begin{align}
  \Pi = \Pi_{\mathrm{sp}} + \Pi_{\mathrm{coh}}. \label{eq:17}
\end{align}
The incoherent (spontaneous) flux $\Pi_{\mathrm{sp}}$ is given by the
part of the sum when the summation indices coincide, i.e., $a =
b$. After integration of this part over the variable
$(\vec k_{\bot} - \vec \theta_a k_z) \Rightarrow \vec k_{\bot}$ the
standard expression of the spectral density of pseudo-photons for the
homogeneous electron beam is obtained \cite{AkhiezerB1965quantum}
\begin{align}
  \Pi_{\mathrm{sp}}
  &= \frac{e_0^2}{v 2\pi^2}N\int d\vec k  \frac{  k^2_{\bot}
  }{[ k_x^2  \gamma^{-2}   + \vec k_{\bot}^2]^2} = \int \omega
  n_{\mathrm{sp}}(\omega) d \omega; \nonumber
  \\
  n_{\mathrm{sp}}(\omega)
  &= N \frac{2 e_0^2}{\pi \omega} \ln\frac{m \gamma}{\omega}, \label{eq:18}
\end{align}
where $|\vec k_{\bot} - \vec \theta_a| \leq \omega
\gamma^{-1}$ and $N$ is the total number of electrons in the beam.

The coherent part is given via the following expression
\begin{align}
  \Pi_{\mathrm{coh}}
  &= \frac{e_0^2}{2 v\pi^2} \int d\vec k  |\vec F(\vec k)|^2, \nonumber
  \\
  \vec F(\vec k)
  &= \sum_a\frac{(\vec k_{\bot} - \vec \theta_a k_x) }{k_x^2
    \gamma^{-2} + (\vec k_{\bot} - k_x\vec \theta_a)^2}e^{- i \vec
    k \vec r_a} e^{i x \vec k  \vec v'_a }, \label{eq:19}
\end{align}
where $\vec v'_a = v\vec \theta_a$. To compute the form factor
$\vec F(\vec k)$ of the beam we need to average the obtained
expression over the distribution on the coordinates $\vec r_{a}$ and
the angles $\vec \theta_{a}$ of the electrons in the beam. For this we
can employ the theory of SASE mechanism of the XFEL, which yields the
following expression for the desired distribution
\cite{XFELTheory,RevModPhys.91.035003}
\begin{align}
  \rho (\vec \theta)
  &= \frac{1}{\pi \sigma^2_{a}} e^{- (\theta_z^2
  + \theta_y^2)/\sigma_{a}^2}, \label{eq:20}
  \\
  f(\vec r)
  &=\frac{1}{\pi \sigma^2_{b}} e^{- ( z^2 + y^2)/\sigma_{b}^2}
    \frac{1}{K}\sum_{l=0}^K\frac{1}{\sqrt{\pi}
    \sigma_{c}}e^{- ( x - l d_0)^2/\sigma_{c}^2}. \label{eq:21}
\end{align}
Here the quantity $\sigma_{a}$ defines the angular spread of the
direction of the velocity, $\sigma_{b}$ is the variance of the
distribution over the transverse coordinates, $d_0$ is the period of
the oscillations of the modulated bunch of length $L_b = K d_0$. It is
also assumed that the parameter $\sigma_{c}$ that defines the
fluctuations of the period of the oscillations $\sigma_{c} \ll d_0$
and the number of the micro-bunches $K \gg 1$. In addition, the
distribution functions are normalized to unity.

Having this distribution, we can approximately substitute the
summation over the discrete index $a$ by the integration over the
continuous variables
\begin{equation}
  \sum_a \Rightarrow N \int d \vec r d \vec \theta f(\vec r) \rho (\vec
  \theta). \label{eq:22}
\end{equation}

Let us now compute the integrals over the coordinates and angles. The
coordinate part is simple and is given via the Fourier transform of
the Gaussian integral
\begin{align}
  \int d \vec r f(\vec r)e^{- i \vec k \vec r}
  &= e^{- ( k_z^2 +  k_y^2)\sigma_{b}^2/4}
    \frac{1}{K}\sum_{l=0}^K e^{-ik_x l d_0}
    e^{- k_x^2\sigma_{c}^2/4} \nonumber
  \\
  &\approx e^{- ( k_z^2 +  k_y^2) \sigma_{b}^2/4}
    \frac{1 - e^{i L_b k_x}}{K(1 - e^{i d_0 k_x})}
    e^{ - k_x^2 \sigma_{c}^2/4}. \label{eq:23}
\end{align}
The averaging over the angular spreads is reduced to the following
integral
\begin{align}
  I = \int d \vec \theta \frac{e^{- \vec \theta^2 /\sigma_{a}^2}}{k_x}
  \frac{1}{\pi \sigma^2_{a}}
  \frac{(\vec \theta_k  - \vec \theta)
  e^{i x k_x v \vec \theta_k \cdot \vec \theta }}
  {(\gamma^{-2} + (\vec \theta_k - \vec \theta)^2)}, \label{eq:24}
\end{align}
where $\vec \theta_k = \vec k_{\bot}/k$. To compute this integral, we
first note that the characteristic angular spread of pseudo-photons is
determined by the parameter $\theta_k \approx
\gamma^{-1}$. Consequently, if the condition $\theta \approx
\sigma_{a} \ll \gamma^{-1}$ is fulfilled we can ignore the influence
of the angular spread of electron on the angular spread of
pseudo-photons. This condition can be fulfilled for realistic
emittances of electron beams \cite{XFELparameters}. As a result, in
this approximation the desired integral is given by
\begin{align}
  I = \frac{1}{k_x}\frac{ \vec \theta_k}{(\gamma^{-2} + \vec
  \theta_k^2)}e^{- (x k v)^2\sigma_{a}^2\vec\theta_k^2/4}. \label{eq:25}
\end{align}
As a result, we can find the expression for the coherent part
$\Pi_{\mathrm{coh}}$ of the pseudo-photons flux
\begin{align}
  \Pi_{\mathrm{coh}}
  &= \frac{N^2 e_0^2}{2 v\pi^2} \int_0^{\infty} d k_x \int d
    \vec \theta_k
    \frac{
    \vec\theta_k^2 e^{-(x k v)^2\sigma_{a}^2\vec\theta_k^2/2 - \vec\theta_k^2 k^2 \sigma_{b}^2/2}
    }
    {( \gamma^{-2} + \vec \theta_k^2)^2  }
    \nonumber
  \\
  &\times\left|\frac{1 - e^{i L_b k_x}}{K (1 - e^{i d_0 k_x}) }\right|^2
    e^{- k_z^2 \sigma_{c}^2/2}\label{eq:26}
\end{align}

We first evaluate the integral over the angles
\begin{align}
  J =  \int d \vec \theta_k
  \frac{\vec \theta_k^2}{( \gamma^{-2} + \vec \theta_k^2)^2}
  e^{- a^2\vec\theta_k^2 }, \label{eq:27}
\end{align}
with $a^2  = 1/2[(xkv)^{2}\sigma_{a}^{2} + k^{2} \sigma_{b}^{2}]$. The
evaluation of this integral is done in the following way
\begin{align}
  J
  &= \pi \int_0^{\infty}du \frac{ u }{(   \gamma^{-2}   +  u)^2  }
  e^{- a^2 u } \nonumber
  \\
  &= \pi \left[\int_0^{\infty}du \frac{e^{- a^2 u }}{(\gamma^{-2} +  u)}
     - \int_0^{\infty}dx \frac{\gamma^{-2}e^{- a^2 u }}{(\gamma^{-2} +  u)^2}
    \right] \nonumber
  \\
  &= \pi \left[-e^{ a^2 \gamma^{-2}}\Ei(- a^2\gamma^{-2}) ( 1+\gamma^{-2}
    a^2) - 1\right]. \label{eq:28}
\end{align}
In this equation $\Ei(x)$ is the integral exponential function
\cite{janke1960losch}.

Thus the coherent part of the spectral density of the
pseudo-photons is represented in the following way ($k_x = \omega/v$)
\begin{align}
  n_{\mathrm{coh}}(\omega)
  &\approx \frac{ N^2e_0^2}{2  \pi \omega v^2}
  \frac{d^2}{L_b^2}[-e^{ a^2 \gamma^{-2}}\Ei(- a^2 \gamma^{-2}) (
  1+\gamma^{-2} a^2) - 1] \nonumber
  \\
  &\times \left|\frac{1 - e^{i L_b \omega/v}}{(1 - e^{i d_0
  \omega/v}) }\right|^2 \exp [ - \omega^2 \sigma_{c}^2/2]. \label{eq:29}
\end{align}

\begin{figure*}[t]
  \includegraphics[scale=0.7]{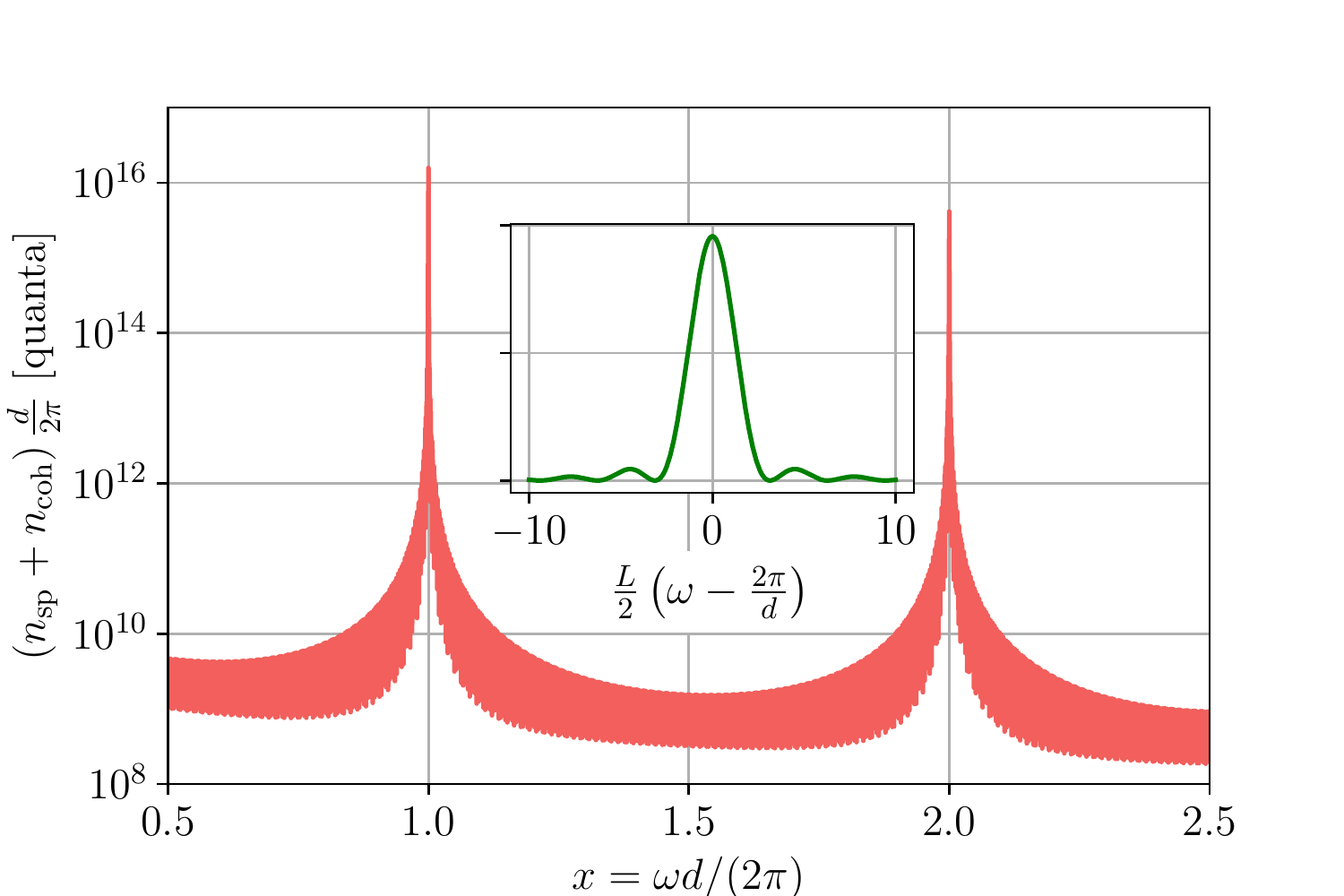}
  \caption{The incoherent and two harmonics of the coherent
    pseudo-photons spectral densities. The inset demonstrates the
    zoomed-in peak structure.}
\label{fig:3}
\end{figure*}

The spectral density $n_{\mathrm{coh}}(\omega)$ of coherent
pseudo-photons has sharp maximums when the frequency
$\omega = 2\pi l/d$, $l = 1,2,...$ and the result can be represented
in the following form
\begin{align}
   n_{\mathrm{coh}}(\omega)
  &\approx \frac{ N^2e_0^2}{2   \pi \omega v^2}
  \frac{d^2}{L_b^2}[-e^{ a^2 \gamma^{-2}}\Ei(- a^2 \gamma^{-2}) (
  1+\gamma^{-2} a^2) - 1] \nonumber
  \\
  &\times \sum_l \frac{\sin^2 [ L_b(\omega - 2 \pi l/d_0)/2v]}{\sin^2 [d(\omega - 2
    \pi l /d)/2v]}e^{ - 2\pi^2 l^2 \sigma_{c}^2/d_0^2}. \label{eq:30}
\end{align}

The multiple $e^{ - 2\pi^2 \sigma_{c}^2/d^2}\leq 1$ as we supposed
that fluctuation of the modulation period defined by the parameter
$\sigma_c < d_0/2 \pi$ .

Let us compare the contributions from the coherent
$n_{\mathrm{coh}}(\omega)$ and incoherent $n_{\mathrm{sp}}(\omega)$
parts of the spectral densities in Eq.~(\ref{eq:18}) . For this we
choose the distribution of the pseudo-photons of the LCLS XFEL
\cite{XFELparameters} facility. The typical electron energy is
$E = 6.7$ GeV which corresponds to the electron gamma factor of
$\gamma \approx 13111$. The parameters $\sigma_{a} = 10^{-4} $ and
$\sigma_{b} = 2\cdot10^{-5}$ cm, parameter
$a^2 \gamma^{-2} \approx 0.2$.  Let bunch the bunch charge be
$Q = 0.2 \ nC$ that corresponds to $N = 1.2 \cdot 10^9$ electrons.  The
duration of the photon pulse we can choose to be 25 fs, which
corresponds to the modulated bunch length of
$L_b = 8.3 \cdot10^{-5}$~cm and the period of the modulation
$d_0 = 10^{-8}$ cm with the parameter $\sigma_c = 10^{-8}$ cm.  In
Fig.~~\ref{fig:3} we plot the incoherent and two harmonics of the
coherent spectral densities of the pseudo-photons of modulated beam
with these parameters.

Finally, the typical frequency spread for the XFEL pulse
$\Delta\omega / \omega \approx 10^{-3}$ and the frequency
$\omega = 6.28 \cdot 10^{8}\ \mathrm{cm}^{-1}$. We can evaluate the
number of the incoherent pseudo-photons in this interval
\begin{align}
  N_{\mathrm{sp}} = n_{\mathrm{sp}}(\omega) \Delta\omega = N \frac{2
  e_0^2}{\pi} \frac{\Delta\omega}{\omega} \ln\frac{m \gamma}{\omega}
  = 1.1 \cdot 10^{5}. \label{eq:31}
\end{align}
This number is significantly smaller that the corresponding number
of photons emitted by the XFEL pulse \cite{XFELparameters}.

At the same time, the integration of the coherent distribution in the
vicinity of the resonant frequency $\omega_{l} = 2\pi l/ d_0$ can be
fulfilled by means of the formula
\begin{align}
  \frac{\sin^2 [ L_b(\omega - 2 \pi l/d_0)/2v]}
  {\sin^2 [d(\omega - 2 \pi l /d)/2v]}
  &\approx 2 \pi v \frac{L_b}{d^2} \delta (\omega - 2 \pi l/d_0)
    \label{eq:32}
  \\
  L_b &\gg d. \nonumber
\end{align}

It yields the number of pseudo-photons in the first harmonic,
comparable with the number of photons in the XFEL pulse
\cite{XFELparameters}
\begin{align}
  N_{\mathrm{coh}}
  &= \frac{N^{2} e_0^2 }{ \omega_0 L_b}
  \left( \ln\frac{1}{a^2 \gamma^{-2}} - C - 1\right)
  \approx 5.9 \cdot 10^{11}; \nonumber
  \\
  a^2 \gamma^{-2}
  &\ll 1 . \label{eq:33}
\end{align}
Here $C \approx 0.577$ is the Euler constant. The number of
pseudo-photons, corresponding to the second harmonic is ten times
less.

As was discussed above, the reflection from the crystallographic
planes leads to the conversion of the pseudo-photons into the real
photons and corresponds to the SPXE process. The reflection
coefficient of the pseudo-photons from the crystallographic planes is
a function of the pseudo-photon frequency. Consequently, the intensity
of the SPXE will reach its maximum when the maximum of the spectral
density of the pseudo-photons coincides with the maximum of their
reflection coefficient. If the position of a crystal is chosen as
shown in Fig.~\ref{fig:2} the pseudo-photons which wave vectors
$\vec k$ satisfy Bragg's condition with one of the reciprocal
lattice vectors $\vec g$ of the crystal will have maximal reflection
coefficient, i.e.,
\begin{align}
  2 \vec k_{0}\cdot\vec g + \vec g^{2} = 0, \label{eq:34}
\end{align}
where $\vec k_{0} = \omega_{0} \vec v / v$.

This means that the crystal should be oriented in such a way to the
electron bunch that the angle between the electron velocity and the
reflection plane equals
\begin{align}
  \theta_{\mathrm{B}} = \arcsin\frac{g}{2\omega_{0}} \label{eq:35}
\end{align}
and the SPXE impulse will propagate in the direction of
$\vec k_{0} + \vec g$ under the angle $2\theta_{\mathrm{B}}$ with
respect to the electron velocity, see Fig.~\ref{fig:2}.

\section{Dynamic theory of SPXE}
\label{sec:dynamic-theory-spxe}

The analysis conducted in the previous sections is valid for the
situation when the crystal is thin enough and the diffraction can be
investigated in the framework of the kinematic theory. However, the
intensity of SPXE reaches its maximum value when the electron
propagates the distances in the crystal larger than the corresponding
extinction length and consequently, the field created by the particles
should be investigated in the framework of the dynamical diffraction
theory \cite{authier2001dynamical}.

\begin{figure*}[t]
 \includegraphics[scale=1.9]{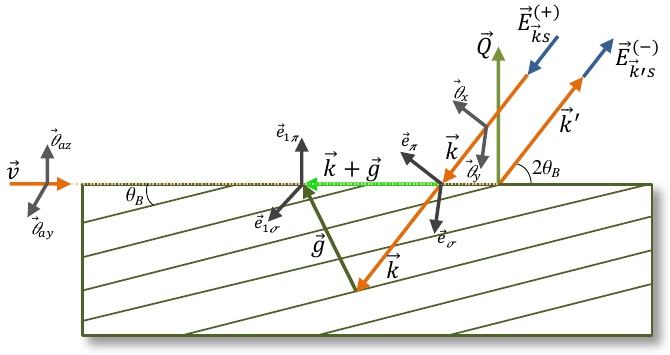}
\caption{The geometry for the dynamical diffraction theory of SPXE}
\label{fig:4}
\end{figure*}

This case for regular PXR was investigated in many works and the
spectral-angular distribution of the emitted number of PXR quanta was
obtained (see \cite{PXR_Book_Feranchuk} and the references there
in). The direct generalization of that results for the case of
modulated beam leads to the following general expression for the
spectral-angular distribution of the emitted number of quanta of SPXE
photons
\begin{align}
  \frac{\partial^2 N_{\vec n \omega s}}{\partial\omega\partial\Omega}
  &= \frac{e_0^2 \omega}{4 \pi^2} \sum_{a}\sum_{b} \int \vec
    E_{\vec k s}^{(+)} (\vec r_a (t),\omega) \vec v_a e^{i\omega t}
    dt \nonumber
  \\
  &\times \int \vec E_{\vec k s}^{(+)*} (\vec r_b (t'),\omega)
    \vec v_b e^{- i\omega t'} dt' , \label{eq:36}
  \\
  \vec r_a (t)
  &= \vec r_a + \vec v_a t, \quad \vec k = - \vec k', \ \vec k' = \omega \vec n \nonumber
\end{align}
where $\vec k'$ is the wave vector of the photons emitted in the solid
angle $d\Omega$, $\vec E_{\vec ks}^{(+)}(\vec r, \omega)$ is the
solution of the Maxwell equations, which describes the diffraction of
the plane wave $\vec e_s e^{i\vec k\cdot\vec r}$ with the polarization
$\vec e_s$ on the crystal and possesses the asymptotic of the plane
wave and an outgoing spherical wave. The use of the wave vector
$\vec k = -\vec k'$ and the field $\vec E_{\vec ks}^{(+)}$ is related
to the fact that for the radiation exited from the crystal one should
exploit the reciprocity theorem of optics \cite{born2013principles}
that relates the waves with different asymptotics, i.e.,
$\vec E_{\vec ks}^{(+)} = \vec E_{-\vec ks}^{(-)*}$. In addition, it
was recently demonstrated that the intensity of the PXR reaches
maximal values in the so called grazing geometry when the extremely
asymmetric diffraction happens. In this case the electrons move along
the crystal surface inside the crystal and the vector
$\vec k_{\vec g} = \vec k + \vec g$ is anti-parallel to the electron
velocity, i.e. $\vec k_{\vec g} \| (-\vec v)$ (see Fig.~\ref{fig:4}).

To find the field created by the particle inside a crystal we can use
the results from the Ref.~\cite{skoromnik_parametric_2019} where a
two-wave approximation \cite{authier2001dynamical, PXR_Book_Feranchuk}
of the dynamical diffraction theory was employed according to which
the field within the crystal is expressed as
\begin{align}
  \vec E_{\vec k s}^{(+)} (\vec r ,\omega)
  &= \vec e_s E_{\vec k s}
  e^{i \vec k \vec r} + \vec e_{1s}E_{\vec k_g s}e^{i\vec k_{\vec g}\vec r},
   \label{eq:37}
  \\
  \vec k_g
  &= \vec k + \vec g. \label{eq:38}
\end{align}
Here $\vec e_{s}$ and $\vec e_{1s}$ ($s = 1,2$) are the polarization
vectors of the incident and the diffracted waves
Fig.~\ref{fig:4}. Their amplitudes satisfy the algebraic system of
homogeneous equations
\begin{equation}
  \label{eq:39}
  \begin{aligned}
    \left(\frac{k^2}{k_0 ^2} -1 - \chi_0 \right) E_{\vec k s}  - c_s
    \chi_{-\vec g} E_{\vec k_g s}
    &= 0,
    \\
    \left(\frac{k_g^2}{k_0^2} -1 - \chi_0 \right) E_{\vec k_g s}  - c_s
    \chi_{\vec g} E_{\vec k s}
    &= 0;
  \end{aligned}
\end{equation}
where $k_0 = \omega$, $\chi_{0}$ and $\chi_{\vec g}$ are the Fourier
components of the crystal susceptibility $\chi(\vec r)$
\begin{align}
  \chi(\vec r)= \sum_{\vec g}\chi_{\vec g} e^{i\vec g\cdot\vec r}.
  \label{eq:40}
\end{align}
The coefficient $c_s = 1 $ for the $\sigma$ polarization ($s=1$) and
$c_s = \cos 2\theta_B$ for the $\pi$ polarization ($s=2$) of the
incident and diffracted waves respectively. In addition, we note that
the waves of different polarizations propagate independently if we
neglect terms of the order of $\sim |\chi_{0}|^{2}$ in the Maxwell
equations \cite{authier2001dynamical,
  benediktovich2013theoretical,SKOROMNIK201786}.

The field amplitudes in vacuum and in crystal should satisfy the
boundary conditions on the crystal-vacuum interface such that the
total field intensity is continuous. Additionally, one needs to take
into account in vacuum not only an incident wave, but also a
specularly reflected diffracted wave
$\vec E_{\vec k_{g}s}^{(sp)} = \vec e_{1s}E_{\vec k_{g}s}^{(sp)}
\exp[i(\vec k_{\|} + \vec g_{\|})\cdot \vec r + i k_{gz}' z]$,
$k_{gz}' = \sqrt{k_{0}^{2} - (\vec k_{\|} + \vec g_{\|})^{2}}$, where
$\vec k_{\|}$ and $\vec g_{\|}$ are the projections of the vectors on
the crystal surface.  As a result the following expressions for the
field inside and outside the crystal were found
\cite{skoromnik_parametric_2019}
\begin{align}
  \vec E_{\vec k s}^{(+)}
  &= \vec e_s e^{i \vec k\cdot\vec r} + \vec e_{1s} E_{  s}^{(sp)}
    e^{i (\vec k_{\|} + \vec g_{\|})\cdot\vec r}
    e^{i k'_{gz} z}, \ z > 0, \label{eq:41}
  \\
  \vec E_{\vec k s}^{(+)}
  &= e^{i \vec k\cdot\vec r}\sum_{\mu = 1,2}
    e^{-i k_0 z \epsilon_{\mu s}}(\vec e_s E_{\mu s}  + \vec e_{1s} E_{ g \mu s}
    e^{i \vec{g}\cdot\vec r}), \ z < 0, \label{eq:42}
\end{align}
where the $x$ axis is directed along the propagation direction of the
electron bunch and the $z$ axis is directed perpendicular to crystal
surface. The coordinate $x$ is changing in the limits $0 < x < L$,
with $L$ being the crystal length, Fig.~\ref{fig:2} and the point
$z = 0$ denotes the crystal surface with $z<0$ is the crystal and
$z>0$ is vacuum. The quantities $\epsilon_{\mu s}$ define the wave
vector inside the crystal and are determined from the condition that
the system of homogeneous equations (\ref{eq:39}) has a nontrivial
solution. The $\vec Q$ is the normal to the crystal surface. The field
amplitudes $E_{s}^{(sp)}$, $E_{\mu s}$ and $E_{g\mu s}$ are found from
the boundary conditions. According to the boundary conditions the
in-plane component $\vec k_{\|}$ of the wave vector is
conserved. Finally, it was found that the main contribution to the
number of emitted photons is given by the amplitude $E_{g1s}$ which
reads
\begin{align}
  E_{ g 1 s}
  &= \frac{c_s \chi_g}{\alpha_B + \chi_0}, \label{eq:43}
  \\
  \alpha_{\mathrm{B}}
  &= \frac{(\vec k  + \vec g)^2 - k_0^2}{k_0^2}. \label{eq:44}
\end{align}
Here the coefficient $\alpha_{\mathrm{B}}$ defines the deviation from the
Bragg's condition.

As in the case of kinematic theory we can split the total intensity in
Eq.~(\ref{eq:36}) into the coherent and incoherent parts. The latter
part corresponds to the situation when in the double sum we select
only the terms with identical indices, i.e., $a = b$. This
contribution is proportional to the number of electrons in the bunch
$N$ and corresponds to the spontaneous parametric X-ray radiation (PXR):
\begin{align}
  \frac{\partial^2 N^{\mathrm{sp}}_{\vec n \omega s}}
  {\partial \omega \partial \Omega}
  &= N  \frac{e_0^2 \omega}{4 \pi^2}
    \left|\int \vec E_{\vec k s}^{(+)} (\vec r (t),\omega)
    \vec v e^{i\omega t} dt \right|^2, \label{eq:45}
  \\
  \vec r  (t)
  &= \vec r_0 + \vec v t. \nonumber
\end{align}

The integral over the electron trajectory was computed in
Ref.~\cite{skoromnik_parametric_2019}, which yields the
\begin{align}
  \frac{\partial^2 N_{\vec n  \omega }^{\mathrm{sp}}}
  {\partial \theta_x  \partial \theta_y}
  &= N \frac{e_0^2}{4 \pi \hbar c } \frac{[(\theta_y -
    \theta_{\mathrm{}ay})^2 + (\theta_x -
    \theta_{\mathrm{a}z})^2 \cos^2 2 \theta_B] |\chi_g|^2 }{[\gamma^{-2} + (\theta_y -
    \theta_{\mathrm{a}y})^2 + (\theta_x -\theta_{\mathrm{a}z})^2 -
    \chi_0']^2}
    \nonumber
  \\
  &\times  \frac{(1 - \ee^{-L k_0\chi_0''|\theta_{\mathrm{a}z}|}
    )}{\chi_0''|\theta_{\mathrm{a}z}|} \ee^{ -\chi_0'' k_0 |z_0| }.
    \label{eq:46}
\end{align}

In this equation the angles $\vec \theta_a, \vec \theta$ - angular
variables of the electron and the emitted photon correspondingly (
Fig.~\ref{fig:4}). They define the small deviations of the electron
velocity the photon wave vector from the ideal values that is
\begin{align}
  \vec v_a = v(1- \frac{\theta_a^2}{2}) \vec e_x + \vec \theta_a;
  \ (\vec e_x \vec \theta_a) = 0; \nonumber
  \\
  \vec k = \vec k_0(1 - \frac{\theta_a^2}{2}) + k_0 \vec \theta;
  \ (\vec k_0 \vec \theta) = 0; \nonumber
  \\
  \omega + (\vec k_0 +\vec g) \cdot \vec v = 0;
  \ 2 (\vec k_0 \vec g) + g^2 = 0. \label{eq:47}
\end{align}

The integral over $\vec \theta$ and averaging over the electron angles
was computed in Ref.~\cite{skoromnik_parametric_2019}. As for example,
for Si crystal the characteristic value of the photon number for PXR
in EAD geometry is
\begin{align}
  N^{\mathrm{sp}}_{\mathrm{PXR}} \approx 2.2 \cdot 10^{-5} \times N
  \approx  0.3 \cdot 10^{5}, \label{eq:48}
\end{align}
for the bunch with charge $Q_0 =0.2$ nC, i.e. $N = 1.3 \cdot 10^{9}$.

This value is comparable with the value of spontaneous pseudo-photons
defined by Eq.~(\ref{eq:31}) that is the crystal is working as the
reflective mirror for this interval of the pseudo-photon spectrum.

Now let us come back to compute the coherent part of the
distribution. If we perform the same substitution from the summation
over the discrete index by the integration over the electron
distribution. In this case the calculation is reduced to the
calculation of the following averaging with the distribution functions
of the bunch
\begin{align}
  \frac{\partial^2 N^{\mathrm{coh}}_{\vec n,\omega s}}
  {\partial \omega \partial \Omega}
  &= N^2  \frac{e_0^2 \omega }{4 \pi^2} |F_s(\vec k)|^2 \label{eq:49}
  \\
  F_s(\vec k)
  &= \frac{1}{N}\int\langle \sum_a
    \vec E_{\vec k s}^{(+)}(\vec r_a (t),\omega)
    \vec v_a e^{i\omega t}\rangle dt, \label{eq:50}
  \\
  \vec v_a
  &= \vec v \left(1 - \frac{\theta_a^2}{2}\right) + v \vec \theta_a,
    \quad \vec v\cdot\vec \theta_a = 0. \label{eq:51}
\end{align}

Here we considered that the electron velocity has deviations from the
$x$ axis and these deviations are described by the $\vec
\theta_{a}$. The vector $\vec v$ is the mean velocity of the electron
bunch and is directed along the $x$ axis ( Fig.~\ref{fig:4}).

As was already discussed the SPXE emission is related to the motion of
the bunch of electrons in a thin layer inside the crystal parallel to
the crystal vacuum interface. In this case the electron velocity is
perpendicular to the normal $\vec Q$ to the crystal surface, i.e.,
$\vec v\cdot \vec Q = 0$. We are interested in the diffracted
wave, that is the wave which is propagating in the direction $\vec
k_{\vec g} = \vec k + \vec g$. Consequently, the main contribution is
given by the $E_{g1s}$ amplitude. Therefore the field which is
entering into the desired averaging reads
\begin{align}
  \vec E_{\vec k s}^{(+)} (\vec r_a (t),\omega) e^{i\omega t}
  &= \vec e_{1s} E_{ g 1 s} \label{eq:52}
  \\
  &\times \exp\{i(\omega + \vec k_g \cdot\vec v
    - \omega \vec Q \cdot\vec \theta \epsilon_{1 s})t \nonumber
  \\
  &\mspace{50mu}+ i v \vec k_g\cdot \vec \theta_a t
    - i\vec k_g\cdot \vec v t\theta_a^2/2
    \nonumber
  \\
  &\mspace{50mu}+ i\vec k_g \cdot\vec r_a
    - i k_0 \vec r_a \cdot\vec Q \epsilon_{1 s} \}. \nonumber
\end{align}

According to the results of Ref.~\cite{skoromnik_parametric_2019} the
SPXE peak is located near the direction defined by the conditions
\begin{itemize}
\item Bragg's diffraction condition
  \begin{align}
    \alpha_{\mathrm{B}} = \frac{(\vec k + \vec g)^{2} -
    k_{0}^{2}}{k_{0}^{2}} = 0. \label{eq:53}
  \end{align}
\item The Cherenkov radiation condition
  \begin{align}
    (\vec k + \vec g)\cdot \vec v = -k_{0} = -\omega. \label{eq:54}
  \end{align}
\item The condition that extremely asymmetric diffraction happens,
  i.e. the diffracted wave is propagating in the direction of
  $\vec k_{g}$, which lies in the crystal plane defined by the normal
  to the crystal surface $\vec Q$
  \begin{align}
    (\vec k + \vec g)\cdot \vec Q = 0. \label{eq:55}
  \end{align}
\end{itemize}

Taking into account these considerations we can represent the vector
$\vec k_{g}$ in the following way
\begin{align}
  \vec k_g = - \frac{\omega}{v^2} \vec v
  \left(1 - \frac{\theta^2}{2}\right) + \omega \vec \theta,
  \quad \ \vec \theta \cdot\vec v = 0, \label{eq:56}
\end{align}

With this the form factor of the bunch is defined by the following
integral with the accuracy $\sim \theta^{3}$
\begin{align}
  F_s(\omega, \vec \theta)
  &= \int_0^L dt \vec e_{1s}\cdot\vec v
    \frac{c_s\chi_g}{\alpha_B + \chi_0}
    e^{i(\omega + \vec k_g \cdot\vec v - \omega \vec Q\cdot\vec \theta \epsilon_{1 s} )t} J,
    \nonumber
  \\
  J
  &= \int d \vec r_a d \vec \theta_a f(\vec r_a)
    \rho(\vec \theta_a)\exp\{i \omega \vec \theta \vec \theta_a t
    + \frac{i\omega t\theta_a^2}{2} \nonumber
  \\
  &\mspace{180mu}+ i\omega(- x_a  +  \vec \theta \vec r_a)\} \label{eq:57}
\end{align}

The evaluation of the integral $J$ over electron angle and coordinates
with the Gaussian distribution Eqs.~(\ref{eq:20})-(\ref{eq:21}) yields
\begin{align}
  J
  &= \exp\left\{
  -\frac{\omega^2 t^2 \theta^2 \sigma_a^2 }
  {4(1 + i\omega t\sigma_a^2)}
  -\frac{\omega^2 \theta^2 \sigma_b^2}{4}-
  \frac{\omega^2  \sigma_{c}^2}{4}
  \right\} \nonumber
  \\
  &\times \frac{1 - e^{i L_b \omega}}
  {K (1 - e^{i d_0 \omega})}. \label{eq:58}
\end{align}

The maxima of $J$ as a function of $\omega$ are located on the
frequencies $\omega$ of the radiation, which are proportional to the
frequencies of the modulation of the electron beam, i.e.,
$\omega = \omega_{0} n$, where $\omega_{0} = 2\pi / d_0$. In addition,
it follows from Eq.~(\ref{eq:58}) that the major contribution to the
coherent impulse of SPXE comes from the part of an electron trajectory
and scattering angles for which the following conditions are fulfilled
\begin{align}
  &\omega t \theta \sigma_{a} < 1, \label{eq:59}
  \\
  &\omega\theta\sigma_{b} < 1. \label{eq:60}
\end{align}

Since the angular spread $\sigma_{a}$ of ultra-relativistic electrons
in the beam of XFEL is significantly smaller than the angular spread
of PXR photons, for which $\theta \simeq \sqrt{|\chi_{0}'|}$
\cite{PXR_Book_Feranchuk} the terms $\omega t \sigma_{a}^{2}$ in
Eq.~(\ref{eq:58}) can be neglected. In this case one can perform the
integration over time $t$ in analytical form. This gives
\begin{align}
  F_s(\omega, \vec \theta)
  &= \frac{c_s(\vec e_{1s}\cdot\vec v)\chi_g}{\alpha_B + \chi_0}
  \frac{1 - e^{i L_b \omega}}{K (1 - e^{i d_0 \omega})} \nonumber
  \\
  &\times e^{-\frac{\omega^2 \sigma_{c}^2}{ 4}-\frac{\omega^2 \theta^2 \sigma_b^2}{4}}
    e^{- \frac{q_s^2}{\delta^2}} \frac{\sqrt{\pi}}{\delta} \nonumber
  \\
  &\times \left[
  \Phi\left(\frac{L\delta}{2} + i\frac{q_s}{\delta}\right)
  - \Phi\left(- i\frac{q_s}{\delta}\right) \label{eq:61}
  \right],
\end{align}
where $\Phi (x)= \frac{2}{\sqrt{\pi}}\int_x^{\infty} e^{- y^2} dy$ is
the complementary error function, $\delta = \omega \theta \sigma_{a}$,
$q_{s} = \omega + \vec k_{g}\cdot\vec v - \omega \vec Q \cdot \vec
\theta \epsilon_{1s}$ and $L_{b}$ is the length of the electron pulse.

According to Ref.~\cite{skoromnik_parametric_2019} when
Eqs.~(\ref{eq:55}-\ref{eq:54}) and the additional condition
$\theta \gg \sigma_{a}$ are satisfied the quantity
$\alpha_{\mathrm{B}}$ is expressed through the angular variables of a
photon as
\begin{align}
  \alpha_{\mathrm{B}} \approx -(\gamma^{-2} + \theta^{2}). \label{eq:62}
\end{align}

As a result, we obtain an expression for the spectral-angular
distribution of photons in the coherent part of SPXE pulse, which for
the first harmonic $u = \omega - 2\pi/d_0$
\begin{align}
  \frac{\partial^2 N^{\mathrm{coh}}_{\vec n \omega s}}
  {\partial \omega \partial \Omega}
  &= N^2  \frac{e_0^2 \omega }{4 \pi^2  }
  \frac{c_s^2 (\vec e_{1s}\cdot \vec v)^2|\chi_g|^2}
  {(\gamma^{-2} + \theta^2 - \chi_0')^2}
  \frac{\sin^2 L_b u/2}{K^2 \sin^2 d_0 u/2} \nonumber
  \\
  &\times \exp\left\{- \frac{2\pi^2  \sigma_{c}^2}{ d^2}
    - \frac{\omega^2 \theta^2 \sigma_b^2}{2}
    - 2 \frac{q_s^2}{\delta^2}\right\} \nonumber
  \\
  &\times \frac{\pi}{\delta^2}
    \left|\Phi\left(\frac{L\delta}{2} - i\frac{q_s}{\delta}\right)
  - \Phi\left(- i\frac{q_s}{\delta}\right)\right|^2. \label{eq:63}
\end{align}

Now we compute the total number of quanta in the coherent part of
SPXE. For this we need to integrate the spectral-angular distribution
(\ref{eq:63}) over the angles and frequencies. To compute the integral
over the frequencies we make use of the following relation
\begin{align}
  \frac{\sin^2 L_b u/2}{K^2 \sin^2 d_0 u/2}
  &\approx 2\pi \frac{1}{L_b} \delta
    (\omega - \omega_0), \label{eq:64}
\end{align}
which is valid when $K = L_b / d_0 \gg 1$. Here $\omega_0 = 2\pi /
d_0$. Consequently, the integration over the frequency becomes
trivial
\begin{align}
  \frac{\partial^2 N^{\mathrm{coh}}_{\vec n \omega s}}
  {\partial \Omega}
  &= N^2  \frac{e_0^2 \omega }{2 L_{b}}
  \frac{c_s^2 (\vec e_{1s}\cdot \vec v)^2|\chi_g|^2}
  {(\gamma^{-2} + \theta^2 - \chi_0')^2} \nonumber
  \\
  &\times \exp\left\{- \frac{2\pi^2  \sigma_{c}^2}{ d^2}
    - \frac{\omega^2 \theta^2 \sigma_b^2}{2}
    - 2 \frac{q_s^2}{\delta^2}\right\} \nonumber
  \\
  &\times \frac{1}{\delta^2}
    \left|\Phi\left(\frac{L\delta}{2} - i\frac{q_s}{\delta}\right)
  - \Phi\left(- i\frac{q_s}{\delta}\right)\right|^2. \label{eq:65}
\end{align}

When an electron bunch moves in the crystal its angular divergence
increases due to the multiple scattering on the atoms of the
crystal. As was demonstrated in Ref. \cite{FImodel} for PXR from
ultrarelativistic electrons, it can be taken into account with the
help of the substitution
\begin{align}
  \gamma^{-2} \Rightarrow
  \tilde{\gamma}^{-2} =  \gamma^{-2} + \theta_s^2,\nonumber
  \\
  \theta_s^2 = \left(\frac{E_s}{E}\right)^2 \frac{L}{L_{R}},
  \label{eq:66}
\end{align}
where $L_{R}$ is the radiation length \cite{ter1972high} and
$E_s \approx 21 \ \mathrm{MeV}$.

Let us now investigate this expression for the case of an ideal
electron beam for which the following conditions are fulfilled
\begin{align}
  \frac{2\pi^2  \sigma_{c}^2}{ d_0^2}
  & < 1, \label{eq:67}
  \\
  \omega^2 \theta^2 \sigma_b^2
  & \approx \omega^2 |\tilde{\gamma}^{-2} - \chi_0'| \sigma_b^2, \label{eq:68}
  \\
  \delta^2 L^2
  & = (\omega \theta \sigma_a)^2 L^2 \approx (\omega L \sigma_a)^2
    |\tilde{\gamma}^{-2} - \chi_0'| < 1. \label{eq:69}
\end{align}

The first condition means that the fluctuations of the modulation
period are rather small. The meaning of the remaining ones is that the
relative transverse width and the angular spread of the particles in
the bunch are less than the angular divergence of the photons in the
PXR pulse \cite{skoromnik_parametric_2019}. These conditions restrict
the exploited electron beams only to the high quality ones with a low
emittance. For example, if we consider electrons with the energy of
14~GeV, propagating through a silicon crystal of a thickness
$L = 1\ \mathrm{cm}$ and generating photons with the energy of 10 KeV
($|\chi_{0}'| \approx 10^{-5}$), then we get the following estimation
for the beam emittance
\begin{align}
  \sigma_{b}^{2}
  &< 10^{-11}\ \mathrm{cm}, \sigma_{a}^{2} < 10^{-9}\
  \mathrm{rad}, \nonumber
  \\
  \epsilon
  &= \sigma_{b}\sigma_{a} < 10^{-10}\
  \mathrm{cm}\times\mathrm{rad}, \label{eq:70}
\end{align}
which is sizable for the LCLS facility.

Under the assumptions of Eqs.~(\ref{eq:67}-\ref{eq:69}) we can further
simplify the expression for the number of emitted photons. For
this we notice that for large $x$ the complementary error function can
be replaced by its asymptotic representation $\Phi(x) \approx
1/\sqrt{\pi} e^{-x^{2}}/x$ and Eq.~(\ref{eq:65}) is further
simplified
\begin{align}
  \frac{\partial  N^{coh}_{\vec n  s}}{\partial \Omega}
  &= N^2  \frac{e_0^2 \omega_0 }{2 L_b}
  \frac{c_s^2 (\vec e_{1s}\cdot \vec v)^2|\chi_g|^2}
  {(\tilde{\gamma}^{-2} + \theta^2 - \chi_0')^2} \nonumber
  \\
  &\times \frac{1}{\pi} \left|
  \frac{1 - e^{i q_s L}}{q_s}\right|^{2}. \label{eq:71}
\end{align}

Now taking into account that when $\omega L \gg 1$ we replace
\cite{PXR_Book_Feranchuk}
\begin{align}
  \frac{1}{\pi} \left| \frac{1 - e^{i q_s L}}{q_s}\right|^{2} =  \frac{4}{\pi}\frac{\sin^2 L q_s/2}{q^2_s}  \approx
  2L \delta(q_{s}) \label{eq:72}
\end{align}
  and integrate over the angle $\theta_{x}$ in the
plane, defined by the vectors $\vec v$ and $\vec g$  with the value
\begin{align}
  q_s
  &= \omega_0 + (\vec k_g \vec v) - \omega_0 \epsilon_{1s}\theta_x
    \nonumber
  \\
  &\approx \omega_0 + (\vec k_0 + \vec g)\cdot  \vec v )
    - \omega_0 \theta_x \sin 2\theta_B = 0; \nonumber
  \\
  \theta_x
  &= \frac{(\vec k_0 + \vec g)\cdot  \vec v}
          {\omega_0 \sin 2\theta_B} = 0. \label{eq:78}
\end{align}

As a result, the photons are emitted with the polarization
proportional to $\theta_y$ and directed along the vector
$\vec v\times \vec g$ ( Fig.~\ref{fig:4}). Thus we get
\begin{align}
  \frac{\partial  N^{\mathrm{coh}}_{\theta_y }}
  {\partial \theta_y}
  &= N^2  \frac{e_0^2 |\chi_g|^2 \theta_y^2 }
    {(\tilde{\gamma}^{-2} + \theta_y^2 - \chi_0')^2}\frac{L}{L_b \sin 2 \theta_B}.
    \label{eq:73}
\end{align}

Finally, after the integration over the remaining angle $\theta_{y}$
one obtains the total number of the SPXE photons:
\begin{align}
  N_{\mathrm{SPXE}} = N^2
  \frac{e_0^2 |\chi_g|^2 \pi}
       {2 \sqrt{\tilde{\gamma}^{-2}  - \chi_0'}}
  \frac{L}{L_b\sin 2 \theta_B}. \label{eq:74}
\end{align}

This result provides a clear qualitative interpretation if we relate
$N_{\mathrm{SPXE}}$ with the number of the coherent pseudo-photons
$N_{\mathrm{coh}}$, defined by Eq.~(\ref{eq:33}). With the logarithmic
accuracy one obtains
\begin{align}
  N_{\mathrm{SPXE}} \approx \frac{\pi}{2 \sin 2 \theta_B}
     \frac{|\chi_g|^2 \omega_0 L}
              {\sqrt{\tilde{\gamma}^{-2}  - \chi_0'}} N_{\mathrm{coh}}.
  \label{eq:75}
\end{align}

The refraction coefficient $R$ of X-ray radiation on the crystallographic
planes is defined by the equation
\cite{benediktovich2013theoretical}$$
R\approx |\chi_g|^2 ( \omega_0 L)^2
$$
Here the Bragg's condition is fullfilled.

If one takes into acccount the ratio of the angular width of the
Bgragg's peak $\Delta \theta_{\mathrm{B}} \approx ( \omega_0 L)^{-1}$
to the angular spread of the pseudo-photons
$\Delta \theta_{\mathrm{ps}} \approx \sqrt{\tilde{\gamma}^{-2} -
  \chi_0'}$ then Eq.~(\ref{eq:75}) takes the following form
\begin{align}
  N_{\mathrm{SPXE}} \approx \frac{\pi}{2 \sin 2 \theta_{\mathrm{B}}}R
     \frac{ \Delta \theta_{\mathrm{B}} }
              {\Delta \theta_{\mathrm{ps}}} N_{\mathrm{coh}}.
  \label{eq:76}
\end{align}

This relation demonstrates that the SPXE emerges as a result of the
reflection of coherent pseudo-photons from the crystallographic
planes. The frequencies of these pseudo-photons are located near the
frequency of the resonance $\omega_{0} = 2\pi / d_{0}$.

The real photons are emitted under the large angle
$2\theta_{\mathrm{B}}$ to the electron velocity. Thus, by choosing
orientation of the crystal the photons can be directed in any desired
location. This allows one to obtain additional experimental windows in
the XFEL experiments.

In order to get a quantitative estimation we consider that an electron
bunch has a charge of 0.2 \ nC and the length
$L_{b} = 8.3 \cdot 10^{-5}\ \mathrm{cm}$.

For the crystal parameters we will employ the values taken from the
X-ray database \cite{StepanovXrayWebServer} for the SPXE radiation
generated in a Si crystal by the reflection (400) and
$\theta_{\mathrm{B}} = \pi/4$
\begin{align}
  \hbar \omega_B
  &= 6.45 \ \mathrm{keV}; \ k_0 = 3.27 \cdot 10^8 \ \mathrm{cm}^{-1};
    \nonumber
  \\ E
  &= 6.7 \ \mathrm{GeV}; \ |\chi_g| = 0.12 \cdot 10^{-4};
    \ \chi_0' = -0.24 \cdot 10^{-5}; \nonumber
  \\
  L
  &= 1.0 \ \mathrm{cm}; \ \tilde{\gamma}^{-2}= 7.5 \cdot 10^{-7}
    \label{eq:76}
\end{align}
The crystal surface is defined by the plane <110>. The photons will be
emitted under the angle $\pi/2$ to the electron velocity.

As a result, we obtai1n the number of photons in the coherent part of
SPXE
\begin{align}
  N_{\mathrm{SPXE}} \approx 5.7 \cdot 10^{12}\ \mathrm{quanta} \label{eq:77}
\end{align}
with the angular spread $\delta\theta$ and spectral width
$\Delta\omega/\omega$ of the order
$\sqrt{\tilde{\gamma}^{-2} - \chi_{0}'|} \approx 0.51\times 10^{-3}$,
which are comparable with the XFEL values.

Also we pay attention to the fact that according to Eq.~(\ref{eq:2})
the higher harmonics of SPXE will be generated (for the harmonic $n$
the condition $\pi n \sigma_{c} < d$ should be satisfied). Moreover
they all will be directed under different angles with respect to the
electron velocity. For example, for the same crystal parameters the
photons with energy $\hbar \omega_B = 12.9 \ \mathrm{keV}$ will be
generated under the angle $40.5^{\circ}$ to the electron
velocity. This allows one to obtain also intensive pulses of the
harder X-rays without the need to change the electron energy that is
not possible for XFEL.

\section{Conclusions}

In the present paper a new application of the modulated electron
bunches is considered. It is supposed that such bunches are formed in
the XFEL undulator due to SASE mechanism and it is shown that the
spectrum of the self electromagnetic field (pseudo-photon spectrum) of
such bunches is essentially transforms and includes intense peaks at
the frequencies proportional to the modulation frequency. When such
bunch goes through the crystal disposed at the undulator exit the
superradiant parametric X-ray emission (SPXE) is generated. Electrons
move in the thin layer along the crystal surface and generate X-ray
pulse that is emitted at the large angle to the direction of the
electron velocity. Intensity of this pulse is proportional to the
square of the number of electrons in the bunch and its characteristics
are comparable with the parameter of the main XFEL pulse, which is
directed along the electron velocity. SPXE pulses can be used for
creation of additional windows for XFEL. Moreover the higher harmonics
of SPXE can be used for generation of the pulses of harder X-rays and
they do not require to change the energy of the electron bunch.

\acknowledgements

The authors are grateful to Prof. A. P. Ulyanenkov for valuable
discussions and support during the project.

\bibliography{pxr_bibliography}

\end{document}